\documentclass[11pt]{amsart}

\usepackage[english]{babel}
\usepackage{bm}
\usepackage{fullpage}
\usepackage{amssymb}
\usepackage{amsmath}
\usepackage{latexsym}
\usepackage{graphicx}

\newtheorem{lemma}{Lemma}

\newtheorem{thm}{Theorem}

\title{Directed paths on a tree: coloring, multicut and kernel}
\author{Olivier Durand de Gevigney}
\author{Fr\'ed\'eric Meunier}
\address{Universit\'e Paris Est, LVMT, ENPC, 6-8 avenue Blaise Pascal, Cit\'e Descartes
Champs-sur-Marne, 77455 Marne-la-Vall\'ee cedex 2, France.}
\email{frederic.meunier@enpc.fr}
\author{Christian Popa}
\author{Julien Reygner}
\author{Ayrin Romero}
\address{Ecole Polytechnique, 91128 Palaiseau Cedex}
\email{\{olivier.durand-de-gevigney,christian.popa,julien.reygner,ayrin.romero-campos\}@polytechnique.edu}

\begin{document}

\begin{abstract}
In a paper published in {\em Journal of Combinatorial Theory, Series
B (1986)}, Monma and Wei propose an extensive study of the
intersection graph of paths on a tree. They explore this notion by
varying the notion of intersection: the paths are respectively
considered to be the sets of their vertices and the sets of their
edges, and the trees may or may not be directed. Their main results
are a characterization of these graphs in term of their clique tree
and a unified recognition algorithm. Related optimization problems
are also presented.

In the present note, we are interested by the case
when the tree is directed, and the paths are considered to
be the set of their arcs. An application is the question of
wavelength assignment in optimal network. In their article, Monma
and Wei show that in this case the intersection graph is a perfect
graph and they give polynomial combinatorial algorithms to solve the
minimum coloring and the maximum stable problems, with the help of
the clique tree representation (Tarjan's method). The maximum clique
problem is immediate. They leave the problem of finding a minimum
clique cover, which is here nothing else than the minimum multicut
of the dipaths, as an open question. Costa, L\'etocart and Roupin in
a survey about multicut and integer multiflow noted that the maximum stable,
which is a maximum set of arc-disjoint dipaths, and the minimum
clique cover, which is the minimum set of arcs intersecting all
dipaths (minimum multicut), are integers solutions of a linear
program with totally unimodular matrices, and hence can be solved
polynomially. For the maximum stable problem (or maximum set of
arc-disjoint dipaths), there is a simple polynomial algorithm found
by Garg, Vazirani and Yannakakis.

In the present paper, we present faster algorithms for solving the
minimum coloring and the minimum clique cover problems (minimum
multicut), and maybe simpler, since we work directly with the
dipaths on the tree and avoid both the clique decomposition and the
linear programming formulation. They both run in $O(np)$ time, where
$n$ is the number of vertices of the tree and $p$ the number of
paths. Another result is a polynomial algorithm computing a kernel
in the intersection graph, when its edges are oriented in a
clique-acyclic way. Indeed, such a kernel exists for any perfect
graph by a theorem of Boros and Gurvich. Such algorithms computing
kernels are known only for few classes of perfect graphs.
\end{abstract}

\maketitle

\section{Introduction}\label{sec:intro}

Consider a directed tree $T=(V,A)$ (with $n$ vertices) and a
collection $\mathcal{P}$ of dipaths in this tree. The intersection
graph is $I(\mathcal{P},T)$ is a graph whose vertex set is
$\mathcal{P}$ and where two vertices are connected by an edge if the
corresponding dipaths share a common arc. Monma et Wei proved that
$I(\mathcal{P},T)$ is then a perfect graph, and explained how such a
graph can be efficiently recognized \cite{MoWe86}. Actually, the
purpose of their paper is an extensive study of the intersection
graph by considering various situations: the tree could be directed
or undirected, and the paths could be identified with their vertex
set or their edge set. Note for instance than when the tree is not
directed, and the path identified with their vertex sets, the
intersection graph is chordal \cite{GyLe70,Ga74,Wa78}, and hence
also perfect.

Once a graph is known to be perfect, min-max formulas hold and
corresponding optimization problems can be solved polynomially, at
least with the ellipsoids (for the general method, see the book of
Gr\"otchel, Lov\'asz and Schrijver \cite{GrLoSc88}). Monma and Wei
gave in their paper combinatorial algorithms for these problems,
except for the minimum clique cover problem. By a theorem of Boros
and Gurvich \cite{BoGu96}, conjectured by Berge and Duchet
\cite{BeDu83}, we know also that if the edges of a perfect graph are oriented
in a clique-acyclic manner, then there is a kernel.

In this paper, we address the problem of finding a minimum coloring,
a minimum clique cover of $I(\mathcal{P},T)$ and a kernel in any
clique-acyclic orientation of the edges of $I(\mathcal{P},T)$. A
coloring of $I(\mathcal{P},T)$ is a color assignment to the dipaths
is such a way that two dipaths sharing a common arc get distinct
colors. The color assignment problem is often interpreted as a
wavelength assignment on an optical fiber (see
\cite{BeCoCoPe06,BeBrCo07} for further studies from this point of
view). A clique cover of $I(\mathcal{P},T)$ is a set of arcs
intersecting all dipaths. Such a set is a multicut of $T$ considered
with the sink and the source of each dipath. A clique-acyclic
orientation of $I(\mathcal{P},T)$ corresponds to a reflexive and
antisymmetric binary relation $\mathcal{R}$ on $\mathcal{P}$
inducing a total order $\preceq_a$ on each subset of dipaths of
$\mathcal{P}$ sharing a common arc $a$. A kernel is a collection
$\mathcal{K}$ of arc-disjoint dipaths of $\mathcal{P}$ such that if
$P\in\mathcal{P}\setminus\mathcal{K}$, then there is
$Q\in\mathcal{K}$ and an arc $a$ such that $Q\succeq_a P$.

\subsection{Definitions}

Let $G=(V,E)$ be a graph. A {\em stable set} of $G$ is a subset
$S\subseteq V$ such that no edge of $G$ connect two vertices of $S$.
The {\em stable set number} is the maximum size of stable set of $G$
and is denoted by $\alpha(G)$.

A {\em clique} of $G$ is a set of vertices any two of which are
adjacent. The {\em clique number} is the maximum size of a clique
and is denoted by $\omega(G)$.

A coloring of $G$ is a color assignment on the vertices of $G$ such
that two vertices connected by an edge get different colors (or
equivalently, a partition of $V$ into stable sets). The minimum
number of colors in a coloring of $G$ is called the {\em coloring
number} and is denoted by $\chi(G)$.

A {\em clique cover} of $G$ is a partition of $V$ into cliques. The
minimum number of cliques in a clique cover of $G$ is called the
{\em clique cover number} of $G$ and is denoted by $\bar{\chi}(G)$.

The following relation are immediate (recall that $\bar{G}$ denotes
the complementary graph of $G$, in which edges become non-edges and
conversely):
$$\alpha(G)=\omega(\bar{G}),\,\, \chi(G)=\bar{\chi}(\bar{G}),\,\,\alpha(G)\leq\bar{\chi}(G),\,\,\omega(G)\leq\chi(G).$$
$G$ is said to be {\em perfect} if $\chi(H)=\omega(H)$ for all
subgraphs $H$ of $G$, or, equivalently, if $\alpha(H)=\bar{\chi}(H)$
for all subgraphs $H$ of $G$, since according to the well-known
theorem of Lov\'asz \cite{Lo72}, $G$ is perfect if and only if
$\bar{G}$ is perfect.

A {\em cycle} of length $s$ in oriented graph is a sequence of arcs
$(v_1,v_2),(v_2,v_3),\ldots,(v_s,v_1)$. A clique-acyclic orientation
of $G$ is an orientation $\vec{G}$ of the edges of $G$ in such a way
that each clique becomes acyclic. Said differently, there is no
cycle of length $3$ in the graph $\vec{G}$.

A {\em kernel} in an oriented graph $D=(V,A)$ is a stable subset
$K\subseteq V$ such that for each $v\in V\setminus K$, there is a
$w\in K$ such that $(v,w)\in A$.

\subsection{The intersection graph $I(\mathcal{P},T)$ as a perfect graph}

\begin{figure}
\includegraphics[width=17cm]{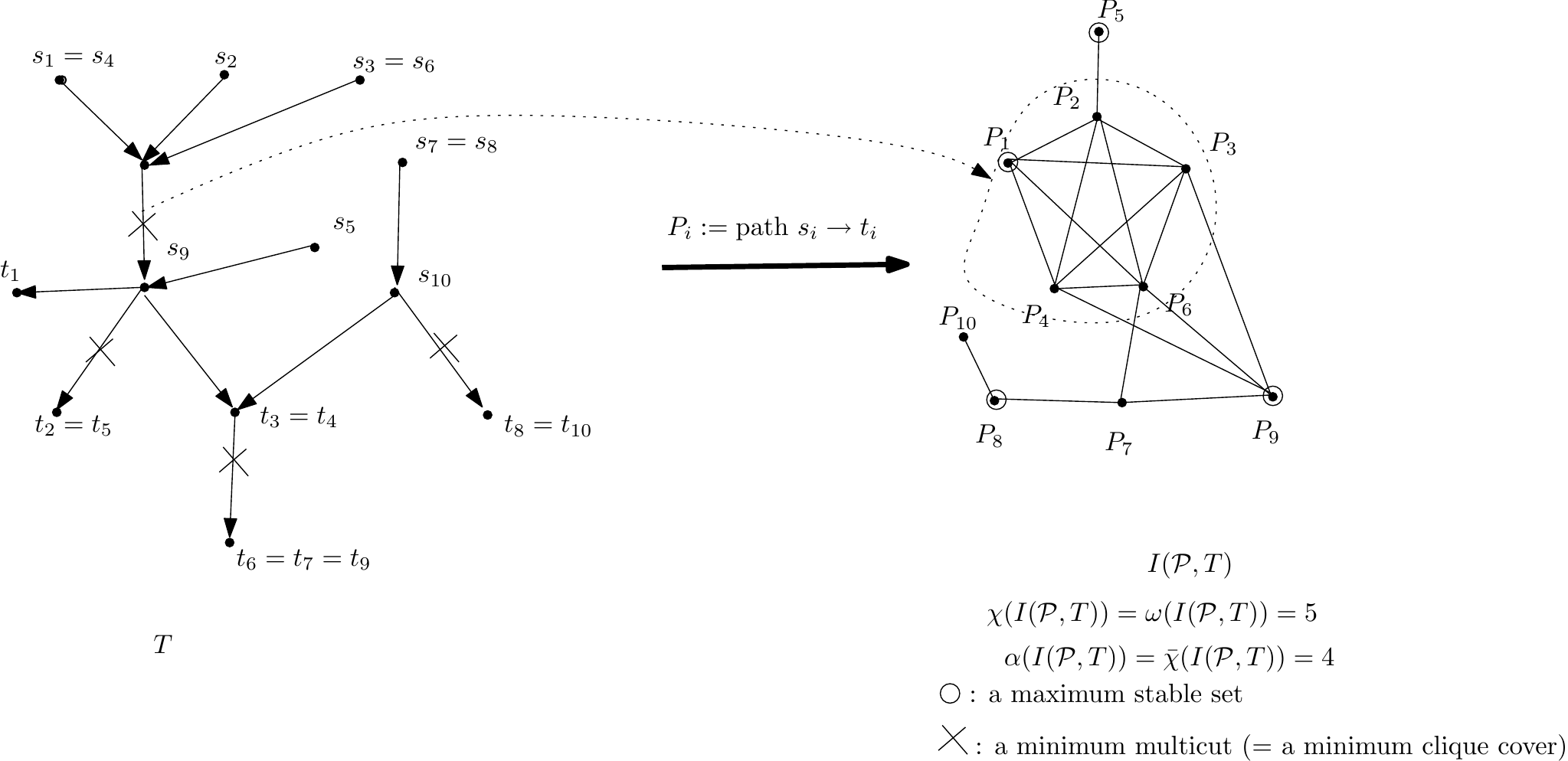}
\caption{Illustration of the perfectness of the
(arc)-intersection graph of dipaths on a directed tree.}
\label{fig:paths}
\end{figure}

Monma and Wei proved that $I(\mathcal{P},T)$ is a perfect graph. In
our context, $\alpha(I(\mathcal{P},T))$ is the maximum number of
arc-disjoint dipaths, $\omega(I(\mathcal{P},T))$ is the maximum
number of dipaths passing through a common arc,
$\chi(I(\mathcal{P},T))$ is the minimum colors that can be assigned
to the dipaths in such a way that two dipaths sharing a common arc
get different colors, and $\bar{\chi}(I(\mathcal{P},T))$ is the
minimum number of arcs intersecting all dipaths of $\mathcal{P}$,
that is a minimum multicut. The equality $\alpha=\bar{\chi}$ says
that the maximum number of arc-disjoint dipaths equals the minimum
multicut, and the equality $\omega=\chi$ says that the maximum
number of dipaths passing through a common arc is equal to the
minimum colors that can be assigned to the dipaths in such a way
that two dipaths sharing a common arc get different colors (see
Figure \ref{fig:paths} for an illustration).

All these value can be computed in polynomial time with the
ellipsoids, as we have said. Monma and Wei gave combinatorial
algorithms for the computation of a minimum coloring and a maximum
stable set of $I(\mathcal{P},T)$. Since they developed a clique
decomposition for their recognition algorithm, a classical method of
Tarjan \cite{Ta85} allows to solve this problem polynomially (and
recursively). For the maximum clique, it is straightforward, just
check each arc one after each other (time complexity $O(n)$). The
minimum clique cover problem was left as an open question. Costa,
L\'etocart and Roupin in their survey about multicut and integer
multiflow \cite{CoLeRo05} noted that the maximum stable set and the
minimum clique cover problems admit a linear programming formulation
with totally unimodular matrices\footnote{actually, network
matrices, but the authors do not seem to have noticed it.}, and
hence can be solved polynomially (for instance with the ellipsoids,
or the interior-points method). For the maximum stable set problem,
the best algorithm is that of Garg, Vazirani and Yannakakis
\cite{GaVaYa97}.

\begin{figure}
\includegraphics[height=8cm]{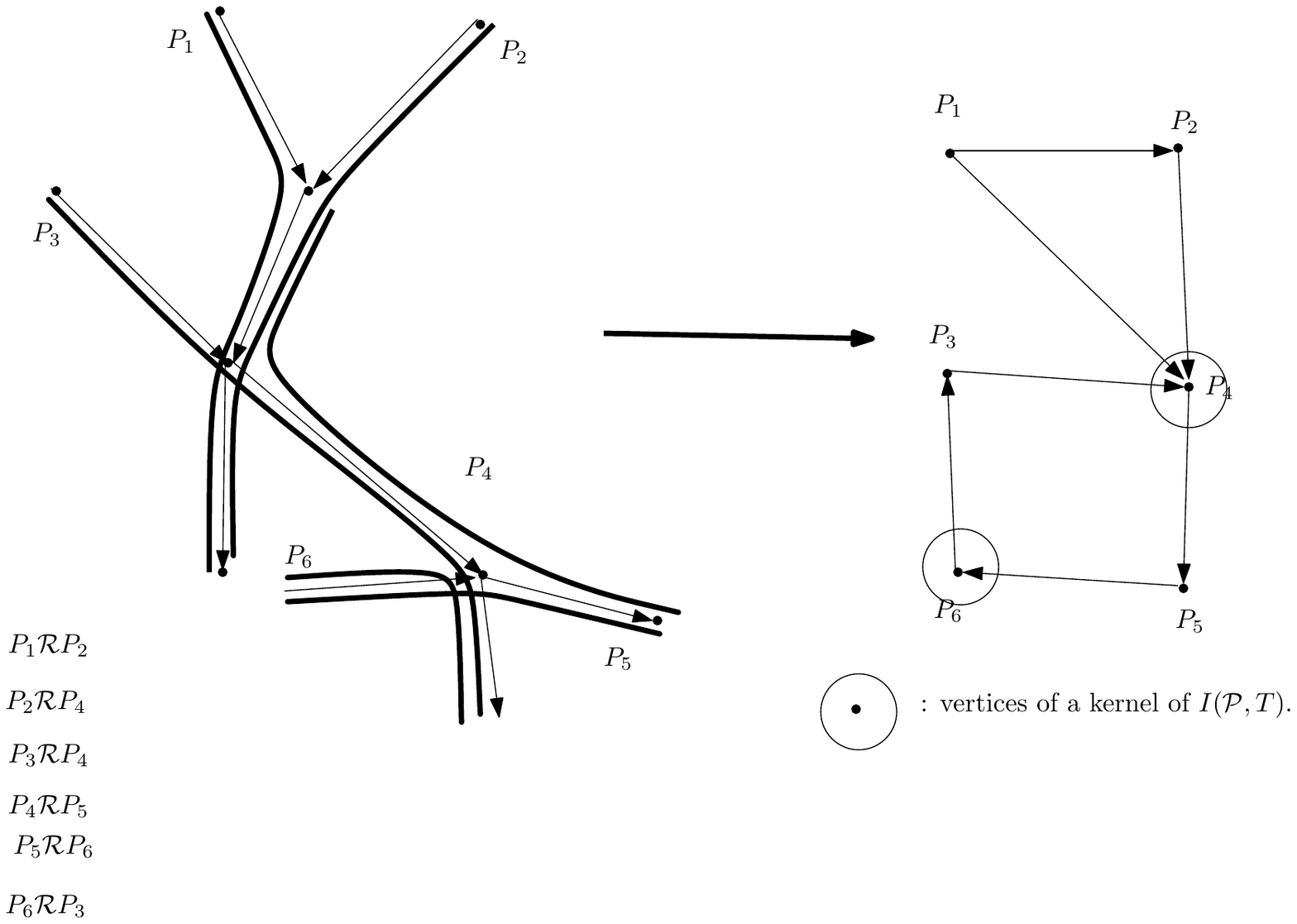}
\caption{A kernel in the arc-intersection graph of the dipaths.}
\label{fig:kernel}
\end{figure}

By the theorem of Boros ang Gurvich already cited, we know that in
any clique-acyclic orientation of $I(\mathcal{P},T)$, there is a
kernel. Any clique-acyclic orientation of $I(\mathcal{P},T)$ induces
a reflexive and antisymmetric binary relation $\mathcal{R}$ on
$\mathcal{P}$ that is a total order $\preceq_a$ when restricted to a
subset of dipaths sharing a common arc $\preceq_a$, and conversely.
See Figure \ref{fig:kernel} for an illustration.

The question whether it is possible to compute in polynomial time a
kernel in a clique-acyclic orientation of any perfect graph is an
open question. Algorithms are known for few classes of perfect
graphs: chordal graphs (straightforward), bipartite graphs
\cite{Ri53}, line graphs of bipartite graphs (as noted by Maffray
\cite{Ma86}, it corresponds to the celebrated stable marriage
theorem, used below, by Gale and Shapley \cite{GaSh62}).

\subsection{Results} In the present work, we show how to find a minimum coloring and a
minimum clique cover of $I(\mathcal{P},T)$ directly on the tree $T$,
neither using clique decompositions nor using linear programming. We
get a time complexity of $O(n|\mathcal{P}|)$ for both the minimum
coloring and the minimum clique cover. It is faster that all
previous known algorithms. We also give an $O(n|\mathcal{P}|^2)$
algorithm that computes a kernel in any clique-acyclic orientation
of $I(\mathcal{P},T)$. It provides a new class of perfect graphs for
which the problem of finding a kernel is known to be polynomial.

\subsection{Plan}

Section \ref{sec:stbip} explains the basic observation on which the
present paper is based. If the tree is a star, all problems can be
reformulated in terms of a bipartite graph: coloring of the path
becomes coloring of the edges, arc-disjoint paths become matching,
and a kernel becomes a stable matching. A tree can be then seen as a
collection of intricate bipartite graphs that
can be processed in a consistent way.

Section \ref{sec:color} is devoted to the minimum coloring problem
(minimum coloring of the dipaths in such a way that two dipaths of
$\mathcal{P}$ sharing at least a common arc get distinct colors),
Section \ref{sec:cliqcover} to the minimum clique cover (minimum
multicut) and Section \ref{sec:kernel} to the kernel.

\section{Starting observation: stars and bipartite graphs}
\label{sec:stbip}

Let $G=(V(G),A(G))$ be an oriented star, and denote by $v$ the
central vertex (a {\em star} is a tree whose diameter is equal to
$2$). Denote by $u_1,\ldots,u_s$ the other vertices. Consider a set
$\mathcal{P}$ of dipaths on this star. We will show that any of the
problems we are interested in can be encoded on a bipartite graph.

\begin{figure}
\includegraphics[width=15cm]{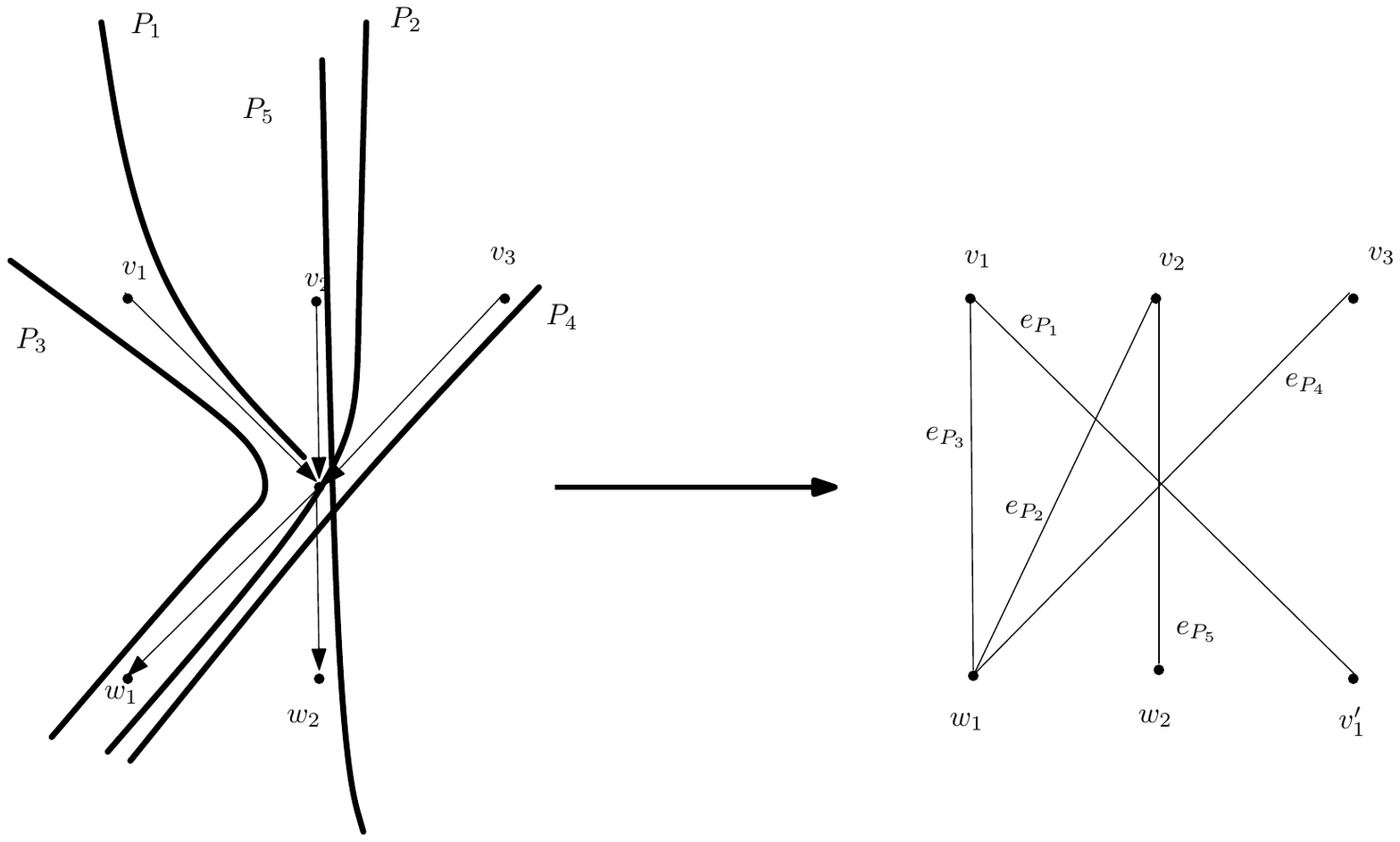}
 \caption{Illustration of the construction of the
bipartite graph that represents the structure of the dipaths passing
through a given vertex.}
\label{fig:bip}
\end{figure}

Adding vertices $u'_1,\ldots$, we can assume that all dipaths have
length $2$. Build the bipartite graph $B$ whose vertices are the
$u_i$ and the $u'_i$ (vertices with outgoing arcs on one side,
vertices with ingoing arcs on the other side) and whose edges
connect vertices in the same dipath (with multiplicities). Hence
there is a one-to-one correspondence between dipaths of $G$ and
edges of $B$. See Figure \ref{fig:bip} for an illustration of
the construction.

Color the dipaths in $G$ is equivalent to color the edges in $B$.
Find a set of arc-disjoint dipaths in $G$ is equivalent to find a
matching in $B$. Finally, find a set of arc-disjoint dipaths in $G$
that provides a kernel in the intersection graph $I(\mathcal{P},G)$
is equivalent to find a stable matching in $B$. Given a graph
$G=(V(G),E(G))$, and a total order $\preceq_v$ on the set of edges
incident to $v$, for each $v\in V(G)$, a matching $M$ is said to be
{\em stable} if for any edge $e\in E(G)\setminus M$, there is an
edge $f\in M$ and a vertex $v$ (incident to both $e$ and $f$) such
that $f\succeq_v e$. 

\medskip

Note that the construction of $B$ can be done in linear time.

\medskip

\noindent{\bf Remark :} The graphs $I(\mathcal{P},T)$ are the common
generalization of the bipartite graphs (see above) and of the
interval graphs (obvious: the underlying tree is then simply a
path).

\section{Minimum coloring}
\label{sec:color}

The strategy to color the dipaths is simple. Choose any
vertex $v$ of $T$. Consider the set of paths
$\mathcal{P}_v$ passing through $v$. Coloring the
paths is equivalent of coloring the edges of a bipartite multigraph
as explained in the previous section. It can
be done in $O(\deg(v)|\mathcal{P}_v|)$, as explained by Schrijver in
his book \cite{Sc03} (Theorem 20.13, p.333). Once these paths are
colored, start again with one of the neighbors of $v$,
consider the set of paths passing through this neighbor $w$, color
the paths of the associated bipartite multigraph without changing
the colors of the paths already colored (those passing through the
arc $vw$), in $O(\deg(w)|\mathcal{P}_w|)$. Start
again, with one of the neighbors of $v$ or $w$, etc. Stop when
all vertices have been processed. The tree structure of $T$ explains
why no color conflict can happen. The whole complexity is then in
$O(n|\mathcal{P}|)$ since $\sum_{v\in V}\deg(v)=2n$.

\begin{thm}\label{thm:color}
Let $T=(V,A)$ be a directed tree and $\mathcal{P}$ a collection of
dipaths in this tree. The minimum number of colors that can be
assigned to the dipaths of $\mathcal{P}$ such that two dipaths
sharing a common arc get distinct color is equal to the maximum
number of dipaths passing through a common arc, and such a minimum
coloring can be found in $O(n|\mathcal{P}|)$, where $n$ is the
number of vertices of $T$.
\end{thm}

The first part of the assertion, which concerns the equality between
the chromatic and clique numbers of $I(\mathcal{P},T)$, was already
proved by Monma and Wei, of course. The algorithm that we propose
below provides actually another (and algorithmic) proof.

\begin{proof}
The proof works exactly as described in the beginning of the
section. For a vertex $v$ of $T$, define $B_v:=(X,E)$ the bipartite
graph that corresponds to $v$ and its neighbors. The dipaths
considered on this star are the dipaths induced by $\mathcal{P}$.

We can color the dipaths while coloring the edges of the bipartite
graphs $B_v$, taking for $v$ the vertices of $T$ one after the
other, with the additional requirement that a new vertex must be the
neighbor of a vertex already processed. The fact that $T$ is a tree
ensures then that when a new bipartite graph $B_v$ is considered,
the edges already colored emanate all from a common vertex (the
vertex already processed). Since the color of the dipaths already
colored is never modified, one keeps an admissible coloring along
the algorithm.

At the end, one has an admissible coloring, and the number of colors
is equal to the maximum degree of the $B_v$, which is exactly the
maximum number of dipaths passing through a common arc. The whole
complexity is $O\left(\sum_{v\in V}\deg(v)|\mathcal{P}_v|\right)\leq
O\left(\sum_{v\in V}\deg(v)|\mathcal{P}|\right)=O(n|\mathcal{P}|)$.
\end{proof}

\section{Minimum multicut}

\label{sec:cliqcover}

A theorem of Gr\" otchel, Lov\'asz and Schrijver explains how to
derive a minimum clique cover in a perfect graph once we know how
to compute a maximum clique and a maximum stable set. This method
would provide an $O(n\times\mbox{DP})$ complexity for our problem,
where $\mbox{DP}$ is the complexity for finding a maximum set of
arc-disjoint dipaths. The best algorithm for this last purpose is
the one proposed by Garg, Vazirani and Yannakakis \cite{GaVaYa97},
and whose complexity is $O(n^{1/2}|\mathcal{P}|)$ at best (the exact
complexity is not easy to determine). Actually, their algorithm is
written for undirected trees, in which case the minimum multicut
problem is NP-hard. Nevertheless, one would get with this approach a
$O(n^{3/2}|\mathcal{P}|)$ time algorithm at best for the minimum
clique cover of $I(\mathcal{P},T)$.

We show in this section how to adapt the algorithm of Garg, Vazirani
and Yannakakis in order to give simultaneously the maximum stable
set and the minimum clique cover of $I(\mathcal{P},T)$, that is a
maximum set of arc-disjoint dipaths and a minimum set of arcs
intersecting all dipaths of $\mathcal{P}$, i.e. a minimum multicut.
Our approach improves the complexity, since the algorithm for the
arc-disjoint dipaths problem is applied only once, and we finally
get a $O(n|\mathcal{P}|)$ time algorithm. Before proving this, we
need to prove a technical lemma about vertex-cover in bipartite
graph.

\begin{lemma}\label{lem:cover}
Let $G=(V,E)$ be a bipartite graph, of color classes $U$ and $W$,
and suppose that there is a vertex $x\in W$ of $G$ contained in
every maximum matching. Suppose given a maximum matching $M$. Then
one can find in linear time a minimum vertex cover $C$ having the
following properties:
\begin{description}
\item[(i)] $x\in C$ and
\item[(ii)] each edge $e$ incident to $x$ and contained in no maximum matching has its other endpoint contained in
$C$, too.
\end{description}
\end{lemma}

\begin{proof}
Let $M$ be a maximum matching. One can assume that there is a
neighbor $y$ of $x$ that is not covered by $M$; indeed, if not, add
a new vertex, call it $y$, and add an edge between $y$ and $x$. One
cannot improve the cardinality of the maximum matching with this new
edge $yx$, since, otherwise, this would mean that $x$ is not in all
maximum matching of $G$. Hence a minimum cover of this new graph is
a minimum vertex cover of the former one.

Make the classical construction of the directed graph $D_M$ (see
\cite{Sc03}) by orienting each edge $e=uw$ of $G$ (with $u\in U$ and
$w\in W$) as follows :
\begin{itemize}
\item if $e\in M$, then orient $e$ from $w$ to $u$.
\item if $e\notin M$, then orient $e$ from $u$ to $w$.
\end{itemize}
Let $U_M$ and $W_M$ be the sets of vertices in $U$ and $W$
(respectively) missed by $M$, and define $R_M$ to be the set of
vertices reachable in $D_M$ from $U_M$. So $R_M\cap W_M=\emptyset$.
Then each edge $uw$ in $M$ is either contained in $R_M$ or disjoint
from $R_M$ (that is $u\in R_M\Leftrightarrow w\in R_M$). Moreover,
no edge of $G$ connects $U\cap R_M$ and $W\setminus R_M$, as no arc
of $D_M$ leaves $R_M$. So $C:=(U\setminus R_M)\cup(W\cap R_M)$ is a
vertex cover of $G$. Since $C$ is disjoint from $U_M\cup W_M$ and
since no edge in $M$ is contained in $C$, one has $|C|\leq |M|$.
Therefore, $C$ is a minimum-size vertex cover and contains $x$.

Now, consider the edge $e$ as in the statement of the lemma. The
other endpoint of $e$ cannot be in $R_M$ since otherwise there is an
$M$-alternating path starting at $U_M$ and containing $e$ -- one
might then switch the edges on this path if it is not starting at a
new vertex $y$ or one might switch the edges on a cycle containing
$e$ and $x$ if it is starting at a new vertex $y$, and in both cases
$e$ would be in a maximum matching of the starting graph. Hence $e$
is necessarily between $U\setminus R_M$ and $x\in W\cap R_M$. Both
endpoints of $e$ are in $C$.
\end{proof}

\begin{thm}
Let $T=(V,A)$ be a directed tree with $n$ vertices and $\mathcal{P}$ be a
collection of dipaths of this tree. The minimum number of arcs intersecting
all dipaths of $\mathcal{P}$ and the maximum number of arc-disjoint
dipaths in $\mathcal{P}$ are equal, and such a minimum set of arcs
can be found in $O(n|\mathcal{P}|)$, where $n$ is the number of
vertices.
\end{thm}

Again, the first part of the assertion, which says that
$\alpha=\bar{\chi}$ for $I(\mathcal{P},T))$ was already proved by
Monma and Wei. Here, we have a new and algorithmic proof.

\begin{figure}
\includegraphics[height=12cm]{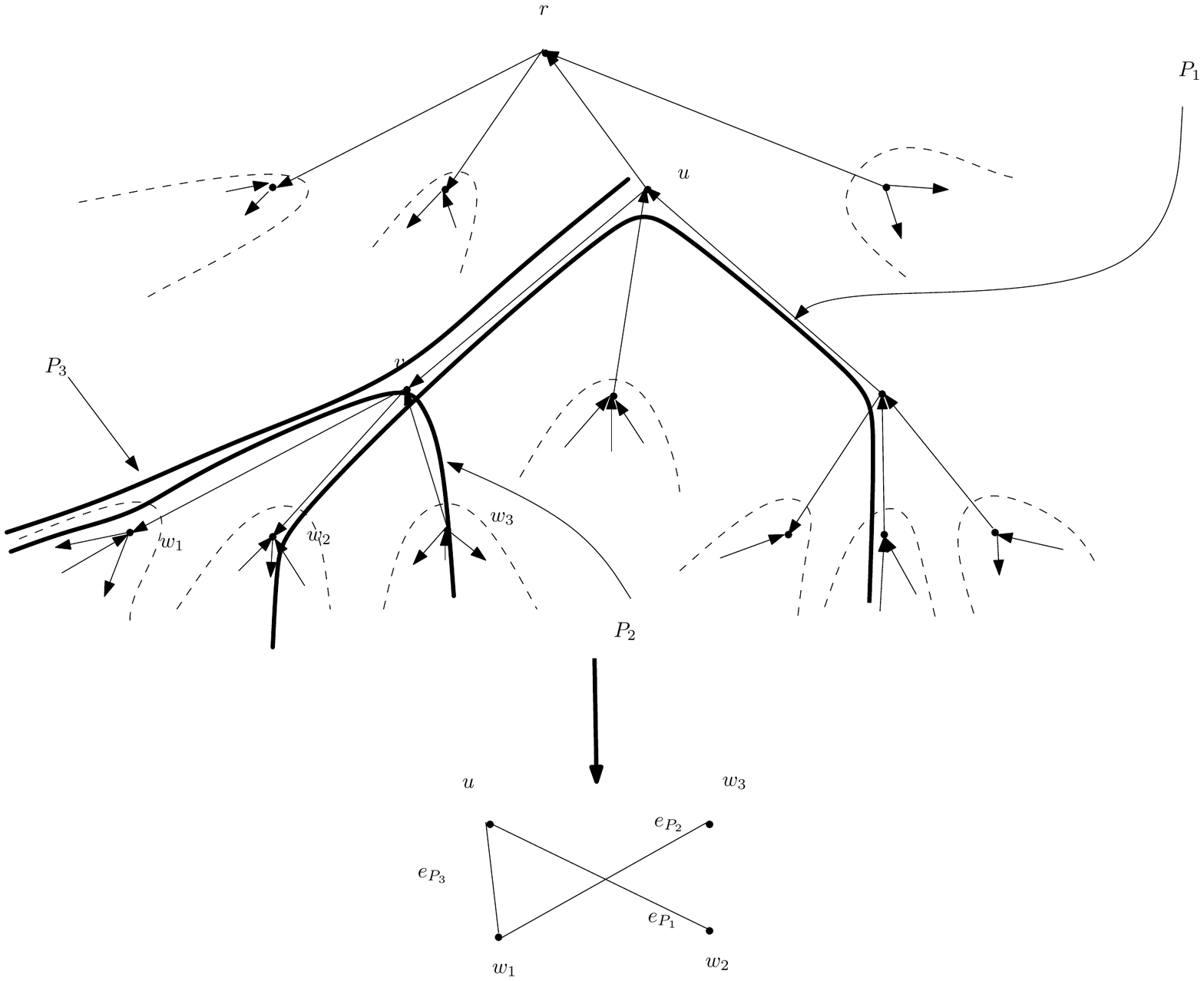}
\caption{We are in the upward pass of the algorithm for multicut. Assume that $P_1\in\mathcal{Q}_{w_2}\setminus\mathcal{B}_{w_2}$, $P_2\in\left(\mathcal{Q}_{w_1}\cup\mathcal{Q}_{w_3}\right)\setminus\left(\mathcal{B}_{w_2}\cup\mathcal{B}_{w_3}\right)$, and $P_3\in\mathcal{Q}_{w_1}\setminus\mathcal{B}_{w_1}$. Then $P_1\in\mathcal{Q}_v\setminus\mathcal{B}_v$ and $P_3\in\mathcal{B}_v$.}
\label{fig:multicut}
\end{figure}

\begin{proof}
The algorithm we describe now follows the scheme proposed by Garg, Vazirani and Yannakakis
\cite{GaVaYa97}.

Root the tree at a vertex $r$. Denote by $\mathcal{Q}_v$
all dipaths that contain the arc $uv$ where $u$ is the father of $v$
in the rooted tree, and by $T_v$ the subtree rooted at $v$. Now
consider the subproblem of finding the maximal number of
arc-disjoint dipaths entirely contained in $T_v$, and denote by
$\alpha_v$ this maximal number. Define $\mathcal{B}_v$ to be the set
of {\em bad dipaths}, that is, those dipaths $P$ of $\mathcal{Q}_v$
such that, when $P$ is selected, the maximum number of arc-disjoint
dipaths entirely contained in $T_v$ and arc-disjoint from $P$ is not $\alpha_v$. Said differently, for
any maximum subset of arc-disjoint dipaths in $T_v$, there is a
dipath in this subset that shares a common arc with $P$. 
By convention, we set $\mathcal{Q}_r:=\emptyset$ and $\mathcal{B}_r:=\emptyset$.

The algorithm works in two pass, an upward one and a downward one.
During the upward pass, all sets $\mathcal{B}_v$ are determined.
During the downward pass, using the sets $\mathcal{B}_v$, a set
$\mathcal{S}$ of arc-disjoint dipaths, and a set $\mathcal{C}$ of
arcs are updated at each vertex $v$. These two sets are such that,
at the end, $|\mathcal{S}|=|\mathcal{C}|$ and each dipath of
$\mathcal{P}$ contains at least an arc of $\mathcal{C}$ (it is the
multicut we are looking for).

\medskip

\noindent{\bf Upward pass: }Process first the leaves. Clearly,
$\mathcal{B}_v=\emptyset$. Then choose a vertex $v$ whose sons have
all been processed. Denote by $u$ its father. Take $P$ in $\mathcal{Q}_v$. If $v$ is an
endpoint of $P$, then clearly $P$ is not in $\mathcal{B}_v$. If $v$
is not an endpoint of $P$, then consider the star $G_v$ induced by
$v$ and its neighbors, and consider on this star the dipath induced
by the non-bad dipaths, i.e. dipaths of $\left(\bigcup_{w\mbox{
\tiny is a son of $v$}}
\mathcal{Q}_w\right)\setminus\left(\bigcup_{w\mbox{ \tiny is a son
of $v$}}\mathcal{B}_w\right)$. According to Section \ref{sec:stbip},
the question of disjoint paths in $G_v$ is in one-to-one
correspondence with the question of matching in a bipartite graph
$B$, in which path $P$ induces an edge $e_P=uw_P$, for some vertex
$w_P$ that is a son of $v$.  Then, $P$ is in $\mathcal{B}_v$ if and
only if there is no maximum matching in $B-u$ that avoids this $w_P$.
The other elements of $\mathcal{B}_v$ are the dipaths of $\mathcal{Q}_v$ that are
already in a $\mathcal{B}_w$, for $w$ a son of $v$. Hence, it is
easy to determine $\mathcal{B}_v$. It can be computed in linear time
once the matching is computed.

\medskip

See Figure \ref{fig:multicut} for an illustration. When the upward
pass is over, $\mathcal{B}_v$ is known for each vertex $v$. We run
now the downward pass.

\medskip

\noindent{\bf Downward pass: }We process first $r$.
Consider the star centered at $r$ and the dipaths induced
by $\left(\bigcup_{w\mbox{ \tiny is a son of $r$}}
\mathcal{Q}_w\right)\setminus\left(\bigcup_{w\mbox{ \tiny is a son
of $r$}}\mathcal{B}_w\right)$. Find a maximum matching $M$ in the
corresponding bipartite graph (Section \ref{sec:stbip}) and a
minimum cover $C$. Define $\mathcal{S}$ to be the dipaths
corresponding to $M$ and $\mathcal{C}$ to be the arcs corresponding
to $C$. We have clearly $|\mathcal{S}|=|\mathcal{C}|$.

Now, take a vertex $v$ whose father $u$ has been processed. Consider
the star centered at $v$ with the dipaths induced by
$\left(\bigcup_{w\mbox{ \tiny is a son of $v$}}
\mathcal{Q}_w\right)\setminus\left(\bigcup_{w\mbox{ \tiny is a son
of $v$}}\mathcal{B}_w\right)\cup\mathcal{Q}_v$. Denote by $B$ the
bipartite graph that corresponds. Two situations may occur:
\begin{itemize}
\item \underline{One of these dipaths has already been fixed in $\mathcal{S}$: } Denote by $f$ the edge in $B$ that corresponds to the already fixed dipath. Find a maximum matching $M$ that uses $f$ and a minimum cover $C$ provided by Lemma \ref{lem:cover}.
Add to $\mathcal{S}$ all dipaths that correspond to edges in $M$. It makes $|M|-1$ new dipaths. Add to $\mathcal{C}$ all arcs $vw$ with $w\in C$ except arc $vu$. It makes $|M|-1=|C|-1$ new arcs. $\mathcal{S}$ and $\mathcal{C}$ are increased by the same quantity. Each dipath in $\left(\bigcup_{w\mbox{ \tiny is a son of $v$}} \mathcal{Q}_w\right)\setminus\left(\bigcup_{w\mbox{ \tiny is a son of $v$}}\mathcal{B}_w\right)$ but not in $\mathcal{Q}_v\setminus\mathcal{B}_v$, contains at least one of these new arcs (according to Lemma \ref{lem:cover}: the vertex $x$ is here played by vertex $u$).

\item\underline{None of these dipaths has already been fixed in $\mathcal{S}$: } Delete from $B$ all edges induced by  the dipaths in $\mathcal{Q}_v\setminus\mathcal{B}_v$. Find a maximum matching $M$ containing no edge induced by a path in $\mathcal{B}_v$ (it is possible by definition of $\mathcal{B}_v$) and a minimum cover $C$ in $B$.
Add to $\mathcal{S}$ all dipaths that correspond to edges in $M$. It makes $|M|$ new dipaths. Add to $\mathcal{C}$ all arcs that correspond to vertices of $C$. It makes $|M|=|C|$ new arcs. $\mathcal{S}$ and $\mathcal{C}$ are increased by the same quantity. Each dipath in $\left(\bigcup_{w\mbox{ \tiny is a son of $v$}} \mathcal{Q}_w\right)\setminus\left(\bigcup_{w\mbox{ \tiny is a son of $v$}}\mathcal{B}_w\right)$ but not in $\mathcal{Q}_v\setminus\mathcal{B}_v$, contains at least one of these new arcs.
\end{itemize}
\medskip

At the end, we get a set of arc-disjoint dipaths $\mathcal{S}$ and a
set of arcs $\mathcal{C}$ of same cardinality. It remains to prove
that each dipath contains at least one arc of $\mathcal{C}$. It is
enough to show that for each dipath $P$, there is a vertex $v$ such
that either
\begin{itemize}
\item we have simultaneously $P\in\mathcal{Q}_w\setminus\mathcal{B}_w$ and $P\in\mathcal{B}_v$ where $w$ is a son of $v$ or
\item we have simultaneously $P\in\mathcal{Q}_w\setminus\mathcal{B}_w$ and $v$ an endpoint of $P$, where $w$ is a son of $v$ or
\item we have simultaneously $P\in\mathcal{Q}_w\setminus\mathcal{B}_w$ and $P\in\mathcal{Q}_{w'}\setminus\mathcal{B}_{w'}$ with $w$ and $w'$ sons of $v$.
\end{itemize}

Indeed we have seen that such dipaths always contain at least one
arc of $\mathcal{C}$. Such a vertex $v$ exists: either there is vertex $v'$ such that $P\in\mathcal{B}_{v'}$, and then take $v$ the farthest vertex of $P$ from $r$ that is such that $P\in\mathcal{B}_v$, or define $v$ to be the nearest vertex of $P$ from $r$.

%choose the farthest vertex $u$ from $r$ such that $P\in\mathcal{B}_u$ and take its (or one of its) son in $P$.

\medskip

The whole complexity is $O(n|\mathcal{P}|)$, since at each vertex
$v$, the complexity is $O(\deg(v)|\mathcal{P}_v|)$ where
$\mathcal{P}_v$ denotes the set of dipaths passing through $v$.
\end{proof}

\section{Kernel}

\label{sec:kernel}

In this section, we assume that we have a reflexive and
antisymmetric binary relation $\mathcal{R}$ defined on the set of
dipaths $\mathcal{P}$ that induces for each arc $a$ a total order
$\preceq_a$ on the set of dipaths containing $a$.

Before stating and proving the main theorem of this section, we
state two theorems about stable matchings in bipartite graphs.

\begin{thm}[Stable marriages theorem, \cite{GaSh62}]
Let $B=(V(B),E(B))$ be a bipartite graph with a total order
$\preceq_v$ on the set of edges incident to $v$ for each vertex
$v\in V(B)$. Then there is stable matching in $B$ and it can be
computed in $O(|V(B)||E(B)|)$.
\end{thm}

\begin{thm}[\cite{McVWi70}]\label{thm:mcv}
Any two distinct stable matchings cover the same set of vertices.
\end{thm}

The algorithm is based on the following lemma, which derives directly from Theorem \ref{thm:mcv}.

\begin{lemma}\label{lem:basic}
Let $B=(V(B),E(B))$ be a bipartite graph with a total order
$\preceq_v$ on the set of edges incident to $v$ for each vertex
$v\in V(B)$. Choose a vertex $u$. Define
$U\subseteq\delta(u)$ to be the set of edges $e\in\delta(u)$ that are in no
stable matching of $\left(B\setminus\delta(u)\right)\cup\{e\}$. Then
for all $f\in\delta(u)\setminus U$ and all $U'\subseteq U$, the edge
$f$ is in all stable matchings of
$\left(B\setminus\delta(u)\right)\cup\{f\}\cup U'$.
\end{lemma}

\begin{proof}
Take $f\in\delta(u)\setminus U$ and $U'\subseteq U$. Let $M$ be a stable matching of $\left(B\setminus\delta(u)\right)\cup\{f\}\cup U'$. 
%Then $u$ is covered by an edge in $M$. Indeed, delete all edges of $U'\setminus M$. The edges of $M$ are still a stable matching of $\left(B\setminus\delta(u)\right)\cup\{f\}$. 
None of the edge $e$ in $U'$ can belong to $M$, otherwise $M$ would be a stable matching of $\left(B\setminus\delta(u)\right)\cup\{e\}$, which contradicts the definition of $U$. Thus $M$ is a stable matching of $\left(B\setminus\delta(u)\right)\cup\{f\}$.
By definition, $f$ is in a
stable matching of $\left(B\setminus\delta(u)\right)\cup\{f\}$, hence there is a stable matching of $\left(B\setminus\delta(u)\right)\cup\{f\}$ for which $u$ is covered. According to Theorem \ref{thm:mcv}, all stable matchings of $\left(B\setminus\delta(u)\right)\cup\{f\}$ cover $u$, and thus, $M$ covers $u$.
As the only edge that can cover $u$ in $M$ is $f$, we have $f\in M$.

%Now, let $e$ be the edge of $M$ covering $u$. Delete all edges of $\delta(u)$ except $e$. The edges of $M$ are still a stable matching of $\left(B\setminus\delta(u)\right)\cup\{e\}$, thus $e=f$, by definiton of $U$.
\end{proof}

We are now in position to state and prove the main theorem of this section.

\begin{thm}
Let $T=(V,A)$ be a directed tree with $n$ vertices and $\mathcal{P}$
be a collection of dipaths of this tree. Assume that there is a
reflexive and antisymmetric binary relation on $\mathcal{P}$
inducing for each arc $a\in A$ a total order $\preceq_a$ on the set
of dipaths containing this arc. Then it is possible to compute in
$O(n|\mathcal{P}|^2)$ a subset $\mathcal{K}$ of arc-disjoint dipaths
of $\mathcal{P}$ such that whenever there is a dipath $Q$ in
$\mathcal{P}\setminus\mathcal{K}$, then there are an arc $a$ and a
dipath $P$ in $\mathcal{K}$ such that $a\in Q$ and $P\succeq_a Q$
(said differently, $\mathcal{K}$ is a kernel of $I(\mathcal{P},T)$).
\end{thm}

The existence of such a $\mathcal{K}$ is already ensured by the
theorem of Boros and Gurvich. Here we have moreover a polynomial
algorithm that computes this kernel.

\begin{proof}
As before, we root the tree at a particular vertex $r$ and
denote by $\mathcal{Q}_v$ all dipaths that contain the arc $uv$
where $u$ is the father of $v$ in the rooted tree, and by $T_v$ the
subtree rooted at $v$.

The algorithm works in the same spirit than the algorithm of
previous section. Instead of having ``bad dipaths'', we will define
{\em uninteresting dipaths} $\mathcal{U}_v$, which are such that,
roughly speaking, when selected, they make impossible the existence
of a kernel in the subtree $T_v$. By convention, we set $\mathcal{Q}_r:=\emptyset$ and $\mathcal{U}_r:=\emptyset$.

As before, the algorithm works in two passes, an upward one and a downward one.

\medskip

\noindent{\bf Upward pass: }If $v$ is a leave, define
$\mathcal{U}_v:=\emptyset$. Now, take a vertex $v$ for which
$\mathcal{U}_w$ has already be defined, for each son $w$ of $v$. Denote by $u$ its father. Let
$B=(V(B),E(B))$ be the bipartite graph that corresponds to the star
centered at $v$ and all its neighbors, with the dipaths induced by
$\left(\bigcup_{w\mbox{ \tiny is a son of $v$}}
\mathcal{Q}_w\right)\setminus\left(\bigcup_{w\mbox{ \tiny is a son
of $v$}}\mathcal{U}_w\right)$. For each $e\in\delta(u)$, test if $e$ is
in a stable matching of $(B\setminus\delta(u))\cup\{e\}$ (use Lemma
\ref{lem:basic}). Then $P$ is in $\mathcal{U}_v$ if the corresponding edge $e$ is not in a stable matching of
$\left(B\setminus\delta(u)\right)\cup\{e\}$. The other elements of $\mathcal{U}_v$ are those dipaths of $\mathcal{Q}_v$ already in a $\mathcal{U}_w$, for $w$ a son of $v$.

\medskip

When the upward pass is finished, we have computed $\mathcal{U}_v$
for all vertices $v$ of the tree $T$.

\medskip

\noindent{\bf Downward pass: } We process first $r$.
Consider the star centered at $r$ and the dipaths induced
by $\left(\bigcup_{w\mbox{ \tiny is a son of $r$}}
\mathcal{Q}_w\right)\setminus\left(\bigcup_{w\mbox{ \tiny is a son
of $r$}}\mathcal{U}_w\right)$. Find a stable matching in
the corresponding bipartite graph. It induces a set $\mathcal{K}$ of
dipaths.

Now, take a vertex $v$ whose father $u$ has been processed. Consider
the star centered at $v$ with the dipaths induced by
$\left(\bigcup_{w\mbox{ \tiny is a son of $v$}}
\mathcal{Q}_w\right)\setminus\left(\bigcup_{w\mbox{ \tiny is a son
of $v$}}\mathcal{U}_w\right)\cup\mathcal{Q}_v$. Denote by $B$ the
bipartite graph that corresponds. Two situations can occur:

\begin{itemize}
\item \underline{One of these dipaths has already been fixed in $\mathcal{K}$: } Denote by $f$ the edge in $B$ that corresponds to the already fixed dipath. Delete from $B$ all other edges that correspond to a dipath in $\mathcal{Q}_v\setminus\mathcal{U}_v$. Compute a stable matching in $B$ (which necessarily contains $f$ by definition of $\mathcal{U}_v$ and according to Lemma \ref{lem:basic}). Add to $\mathcal{K}$ the dipaths whose corresponding edges are in the stable matching.
\item \underline{None of these dipaths has already been fixed in $\mathcal{K}$: }Delete from $B$ all edges that correspond to a dipath in $\mathcal{Q}_v\setminus\mathcal{U}_v$. Compute a stable matching in $B$. Add to $\mathcal{K}$ the dipaths whose corresponding edges are in the stable matching. By definition of $\mathcal{U}_v$, no dipath using arc $uv$ is added to $\mathcal{K}$.
\end{itemize}

At the end, we get a set $\mathcal{K}$ of arc-disjoint dipaths. It
remains to check that whenever a dipath $P$ of $\mathcal{P}$ is not
in $\mathcal{K}$, there is an arc $a\in P$, and a dipath
$Q\in\mathcal{K}$ such that $a\in Q$ and $Q\succeq_a P$. It is
enough to show that for each dipath $P$, there is a vertex $v$ such
that 
either
\begin{itemize}
\item we have simultaneously $P\in\mathcal{Q}_w\setminus\mathcal{U}_w$ and $P\in\mathcal{U}_v$ where $w$ is a son of $v$ or
\item we have simultaneously $P\in\mathcal{Q}_w\setminus\mathcal{U}_w$ and $v$ an endpoint of $P$, where $w$ is a son of $v$ or
\item we have simultaneously $P\in\mathcal{Q}_w\setminus\mathcal{U}_w$ and $P\in\mathcal{Q}_{w'}\setminus\mathcal{U}_{w'}$ where $w$ and $w'$ are sons of $v$.
\end{itemize}

Indeed such dipaths are edges of a bipartite graph for which a
kernel is computed. As in the previous section, such a vertex $v$
exists: either there is vertex $v'$ such that $P\in\mathcal{U}_{v'}$, and then take $v$ the farthest vertex of $P$ from $r$ that is such that $P\in\mathcal{U}_v$, or define $v$ to be the nearest vertex of $P$ from $r$.

%choose the farthest vertex $u$ from $r$ such that $P\in\mathcal{U}_u$ and take its (or one of its) son in $P$.

\medskip

The whole complexity is $O(n|\mathcal{P}|^2)$, since at each vertex
$v$, the complexity is $O(\deg(v)|\mathcal{P}_v|^2)$ where
$\mathcal{P}_v$ denotes the set of dipaths passing through $v$: we
repeat at most $|\mathcal{P}_v|$ times the computation of a stable
matching for each vertex $v$.
\end{proof}

\bigskip

\noindent{\bf Acknowledgement} We are grateful to Michel Cosnard for
having pointed out the question of a simple and fast algorithm for
the coloring problem.

\bibliographystyle{amsplain}
\bibliography{Combinatorics}

\end{document}